\documentclass{article}
\usepackage{iclr2019_conference,times}
\usepackage{hyperref}
\usepackage{url}
\usepackage{graphicx}
\usepackage[font=small,skip=2pt]{caption}

\title{Regulatory Markets for AI Safety}
\author{Jack Clark \thanks{OpenAI}\\ \texttt{jack@openai.com} \\
\And Gillian K. Hadfield \thanks{Schwartz Reisman Institute for Technology and Society, University of Toronto, Vector Institute for AI, University of California Berkeley Center for Human-Compatible AI, and OpenAI. We thank Jan Leike, Frank Partnoy, participants at workshops at the University of Chicago and NBER Economics of AI Conference as well as other colleagues for comments on earlier drafts.} \\ \texttt{g.hadfield@utoronto.ca}}

\iclrfinalcopy
\begin{document}

\maketitle

\begin{abstract}
We propose a new model for regulation to achieve AI safety: global regulatory markets. We first sketch the model in general terms and provide an overview of the costs and benefits of this approach. We then demonstrate how the model might work in practice: responding to the risk of adversarial attacks on AI models employed in commercial drones.

\end{abstract}

\section{Introduction}
Ensuring that humans benefit from AI will depend on our capacity to regulate how it is built and deployed. There are risks arising from misuse of powerful technology and from misalignment between the goals an AI system is pursuing and the true goals of a relevant group of humans. Containing both types of risks requires adapting our existing regulatory systems and building new tools to regulate how AI is built, used, and safeguarded against use by malicious actors. Our existing regulatory systems are, however, struggling to keep up with the challenges posed by a fast-moving technology-driven global economy \citep{Marchant2011,Hadfield2017} and AI only promises further challenges. 

Regulatory strategies developed in the public sector operate on a time scale that is much slower than AI progress, and governments have limited public funds for investing in the regulatory innovation to keep up with the complexity of AI's evolution. AI also operates on a global scale that is misaligned with regulatory regimes organized on the basis of the nation state. This leads to competitive dynamics that manifest at the level of companies and nations: pressure to keep up in the geopolitical sphere sharpens the collective action problem further, with national regulation potentially slowed or foregone in the interests of maintaining competitiveness. 

Examples of the ways in which our existing regulatory regimes are struggling to keep up with even the relatively contained uses of AI we see deployed today abound.  Machine learning algorithms deployed in the criminal justice system, such as the COMPAS algorithms used in U.S. bail hearings and sentencing, have well-known problems with racial bias that courts are poorly equipped to address:  court challenges are costly, many defendants affected by biased algorithms are represented by severely overburdened public defenders with limited resources to investigate and limited to no access to software tools comparable to those available to enforcement,\footnote{https://www.americanbar.org/news/abanews/aba-news-archives/2014/02/public\_defender\_case/; https://www.nytimes.com/2019/11/22/business/law-enforcement-public-defender-technology-gap.html}; trade secret laws and contracts with private software providers often block scrutiny of these algorithms.  The targeted use of machine learning algorithms to flood social networks with “fake news” and malicious bots impersonating real people have contributed to the effort to interfere with elections and undermined public confidence in electoral systems in both the U.S. and Europe.  U.S. Congressional hearings into the Cambridge Analytica scandal, in which massive amounts of personal data were collected to feed AI-powered political influence campaigns, demonstrated how little legislators understood about the problem they were called on to address.  

No good solutions are in sight: our existing systems lack the capacity, for example, to reliably identify fake news without human input, which is infeasible on volumes seen by platforms like Facebook. Google still has not developed a solution to the problem it discovered in its AI-powered photo tagging algorithms, which tagged black people as gorillas, other than to eliminate gorilla from the available labels.  The self-driving car industries still do not have effective regulatory regimes in place that balance the risks of these technologies against their benefits; following a 2018 crash, Uber suspended entirely its deployment of self-driving cars. The 2018 changes to privacy law in Europe--the General Data Protection Regulation--have ambitious goals including the right to have irrelevant or obsolete data removed from search engines and explanations given for AI-based decisions. But governments have few tools to implement these promises. Implementation is still widely dependent on the voluntary or self-regulatory efforts of the companies deploying AI. And indeed, most of the regulatory effort we see is in the realm of self-regulation and industry standard-setting, such as the development of IEEE guidelines on ethical AI. These are laudable efforts but ultimately leave the bulk of the power to determine how AI is, and is not, used in the hands of commercial entities with a conflict of interest.  

Adding to the challenge are competitive race dynamics: Companies competing in markets have an incentive to build AI faster than their competitors, and as the above examples show, assuring the safety of large-scale machine learning-driven systems appears to be both costly and difficult; slowing that process down while encouraging an environment for investment to ensure safe development is a collective action problem that regulation is needed to address.  

Military and national interests are also prompting a competitive environment, with China announcing in 2017 a goal to be a world class leader in AI theory, technology, and application by 2030. This and other related national plans and investment initiatives around the world have ratcheted up tensions in how nations cultivate their own technology sectors, while countering the perceived influences of other countries. These elevated tensions have been most recently dramatized by the United State's actions against Huawei which have included bans on the company contracting with US companies without government approval, causing spillover effects that seem set to lead to Huawei altering its technological supply chain.  Situations like this highlight how pressure to keep up in the geopolitical sphere sharpens the collective action problem further still, with national regulation and international collaboration potentially slowed or foregone in the interests of maintaining national competitiveness. 

We can adapt regulatory systems to keep up with AI and the broader trend of more advanced digital technologies by accelerating the pace and quality of efforts to innovate better ways to regulate a rapidly evolving percentage of the world's R\&D and industrial base. One way to do this is by making public sector regulation more responsive, following the lead of industry approaches to developing goods and services; this is the challenge of “agile governance” \citep{WEF2018}. But public sector solutions are inevitably limited. Increasing the ability to generate innovative regulatory approaches is likely to require moving more of the problem of regulation out of the domain of the public sector and into the domain of markets: creating markets for regulation that attract money and talent to the problem. 

A critical challenge for ensuring that AI follows a safe and beneficial path of development, then, is to find ways to adapt our regulatory systems to the pace, complexity, and global reach of AI and to do so in ways that are simultaneously:
\begin{itemize}	
\item as smart as the AI systems they seek to regulate
\item not dictated by industry actors alone
\item capable of coordinating across nations.
\end{itemize}

In this paper, we propose a new approach to regulation that meets these three challenges: \emph{global regulatory markets}. In this approach, governments create a market for regulation in which private sector organizations compete to achieve regulatory outcomes set by a government regulator. We envisage regulatory markets as being regulated at national and international scale. Any regulatory market is composed of three principal actors:  The targets of regulation, private regulators, and governments. Targets are businesses and other organizations the behavior of which governments seek to regulate. In the AI context, these are companies and public agents that are building and/or deploying AI as products or processes. Private regulators are for-profit and non-profit corporations/organizations that enter into the business of developing and supplying regulatory services which they compete to sell to targets. Governments require targets to purchase regulatory services (entering into a regulatory contract with a private regulator) and directly regulate private regulators and the market for regulatory services, ensuring it operates in the public interest.  

A key distinction between regulatory markets and self-regulation is that in regulatory markets the overall outcomes that must be achieved by regulation are determined by the government and not the regulated entity itself. Governments in this regime design and oversee the structure of the regulatory market and regulate the entities within it.  

Our discussion is organized as follows.  We first review the existing regulatory landscape, identifying the tools, mechanisms, and institutions available to governments seeking to ensure that AI technologies are developed and deployed in accordance with politically-set goals.  We then lay out the details of our model and develop these details in the concrete setting of commercial drone regulation. In subsequent sections we discuss the benefits and risks of this approach to regulation, including the risk of capture and collusion and how these risks might be addressed through market design. We spend some time examining four examples of existing reliance on hybrid public-private regulation. Two of these show the limits of these mechanisms in complex and global settings outside of the AI domain and highlight lessons for improved regulatory design: credit rating agencies in the context of the 2008 global financial crisis and the 2018-2019 Boeing 737-MAX crashes.  Two other examples--the use of private certification providers to oversee medical device quality and the reliance on approved regulators in legal services in the United Kingdom--suggest other more hopeful lessons that can be drawn. We conclude with thoughts on how this model might be implemented initially through voluntary industry efforts as a precursor to government oversight and how a government ready to pilot this system could begin.
\section{The Existing Regulatory Landscape}
Modern societies rely on multiple systems to channel the behavior of people and organizations in ways that are consistent with collectively-determined interests. Markets channel resources towards the production of goods and services that people want (as expressed by demand) and using processes that minimize the cost of production.  Tax and subsidy laws redistribute income so as to ensure that the capacity to register values with demand is more fairly shared. Regulatory agencies develop rules and procedures to improve the performance of markets: protecting contract and property interests, deterring fraud and collusion, and overcoming market failures and externalities. Political systems give people the opportunity to choose their representatives or run for office themselves. Constitutions encode broad commitments to freedom or anti-discrimination and limit the power of governments; courts, constituents, and a free press monitor compliance with those commitments.
\subsection{Methods of Regulation}
There are currently four principal methods for regulating markets and industries \citep{May2007, Carrigan2011}.  The traditional, and most widely used, approach is prescriptive, sometimes called command-and-control.  Prescriptive regulation supplies specific and sometimes highly detailed rules governing behavior, technology, and/or processes; failure to comply with the rules generates penalties (fines, loss of authority to provide goods or services, criminal sanctions, etc.)  Also traditional and widely used is licensing (which can also be thought of as a form of prescriptive regulation and is also called prior approval): the requirement of obtaining and maintaining authorization before providing goods or services in markets. Initial authorization can require completion of prescribed education, testing of individuals or products, inspection and evaluation of facilities or processes, or an evaluation of and plan for managing potential risks \citep{Ogus2004, Kleiner2000}. Maintenance of a valid license can require ongoing compliance with regulations.  Operating without a license is penalized.   

In recent decades, these traditional forms of regulation have been supplemented with “new governance” techniques.  Performance-based regulation (also called outcomes-based or principles-based) specifies results (sometime expressed as metrics, sometimes expressed only as principles) that a provider must achieve but does not specify how the provider is to achieve those results \citep{Coglianese2003,May2011}.  Failure to achieve outcomes is penalized.  Management-based regulation (also called process-oriented, risk-based, or enforced self-regulation) requires firms to evaluate the risks generated by their business and to develop their plan for how those risks will be managed. Plans might need approval from government or a third-party certification agency. Failure to generate a plan as required and/or to abide by the plan is penalized \citep{Gilad2011,Coglianese2003,Braithwaite1981,AyresBraithwaite1992,Braithwaite2011}. Meta-regulation embeds these new governance techniques in a system in which both regulated entities and government regulators continually learn from experience to update required processes and outcomes \citep{Gilad2010}. 

The move to new modes of regulation has been fostered by the perception that traditional approaches inhibit both efficiency and innovation in the achievement of regulatory goals.  The theory of new governance approaches is that government should find ways to harness the expertise and cost-minimizing incentives of industry itself in the pursuit of politically-established outcomes such as a safe food supply, reduced pollution, or stable financial systems.  These are insights on which our model of regulatory markets are premised.
\subsection{Private Regulation}
Although we think of regulation as a state-led activity, private non-state actors have long played a role in regulation and there are many examples of a shift from command-and-control to outcome-based, principles-based or risk-based regulation \citep{Gilad2010}. As emphasized in Hadfield (2017), there is an economic demand for legal infrastructure to make market (and other) interactions more reliable and productive and as a result private entities can in some cases find a profitable opportunity to supply rules to meet that demand; in other settings, private actors can meet the demand for legal infrastructure by collectively establishing industry standards and funding an oversight mechanism. There is a robust literature in political science examining these forms of voluntary ``private regulation'' \citep{Buthe2010}. For example, the International Organization for Standardization (ISO) provides voluntary standards in areas ranging from quality (ISO9000) and the environment to risk management and food safety. Globally, food safety standards are implemented with a broad mix of public standard-setting and private certification and monitoring to ensure compliance \citep{Rouviere2017}. 

Voluntary submission to regulation can also go beyond voluntary compliance, as when voluntarily chosen standards are made enforceable through contract. Publicly-enforceable standards of conduct are privately written into contracts both by parties to the contract, as when an online retailer sets out rules governing privacy and data usage in its terms of service, and by third parties, as when trade associations require their members to use the organization’s contract terms \citep{Bernstein1996}. Global companies  such as Nike, Apple, and Walmart now routinely use their supply chain contracts to impose standards pertaining to subjects such as quality control, environmental practices, workplace safety, and child labor on suppliers in countries with underdeveloped or dysfunctional regulatory systems \citep{Locke2013}. In many cases, supplier compliance with supply contract obligations is monitored and enforced by private sanctions (contract termination, fines) imposed by the purchasing company, which may outsource oversight to a third-party monitor \citep{Short2016}. An example of this is the response of large retailers in the garment industry to the 2013 collapse of the Rana Plaza factory and the 2012 Tazreen Fashions factory fire, both in Bangladesh. American retailers such as the Gap established the Alliance for Bangladesh Worker Safety which created a set of workplace safety standards for members to incorporate into their supply contracts and a monitoring facility to inspect factories for compliance.\footnote{http://www.bangladeshworkersafety.org/} European retailers such as H\&M went beyond the Alliance approach in response to the Bangladeshi disasters and established the Accord on Fire and Building Safety in Bangladesh.\footnote{http://bangladeshaccord.org/} The Accord makes the achievement of workplace safety standards and commitments to fund safety programs in Bangladeshi factories subject to third-party enforcement (arbitration) and is overseen by a governing board that includes union representatives and is chaired by a representative from the UN's International Labour Organization (ILO). 

Even if we limit our purview to mandatory regulation, however, where regulation is imposed by the state, private entities have long been a significant presence.  There are numerous examples of cases in which public regulation has piggybacked on systems initially developed privately\footnote{Securities regulation originated, for example, in the private regimes developed by stock exchanges \citep{Macey1999,Seligman2003,Birdthistle2013}.} and this creates an incentive for industries to organize self-regulation in order to shape what is seen as inevitable public regulation \citep{Ostrom1990, Braithwaite2000}. Private actors also play an indirect role through their influence over government regulation \citep{Haas1992}. Standards developed by private standard-setting bodies--membership organizations such as the Society of Automotive Engineers, for example--are sometimes incorporated into legislation.\footnote{See, e.g., 16 CCR § 3351.6 ``Equipment Requirements for Automotive Air Conditioning Repair Dealers'' (all automotive repair dealers engaged in service or repair of air conditioning systems in vehicles must have refrigerant identification equipment that meets or exceeds Society of Automotive Engineers standard J1771, ``which is hereby incorporated by reference.'')} Privately-developed rules (established and sometimes monitored by industry bodies or by individual firms such as insurers) can also be imposed by government as a condition of obtaining a government contract or permit.  Private membership organizations are sometimes delegated authority to regulate their members on behalf of government actors; examples include FINRA and bar associations.  The demand for transnational regulatory standards in our increasingly integrated global economy has also resulted in increasing reliance on private actors to regulate.  As with domestic regulation, there has long been widespread reliance on international standard-setting bodies to supply the rules governing goods and services sold in global markets \citep{Braithwaite2000, Buthe2011}. 

As we have noted, the pressure to develop ``new governance'' alternatives to conventional command-and-control regulation arises largely as a consequence of the growing need for what is sometimes called ``agile'' governance (World Economic Forum 2016, World Economic Forum 2018) in the face of rapid change and high levels of complexity.  The much-heralded creation of a broad European ``right to be forgotten'' in search results, for example, has been implemented by requiring search engines to themselves serve the function of hearing and ``adjudicating'' claims; in 2016 Google estimated its internal ``quasi-judicial'' panels were evaluating close to 600 claims a day.  This pressure to default back to self-regulation is a response to limited technical and financial resources in governments.\footnote{https://www.nytimes.com/2016/04/19/technology/google-europe-privacy-watchdog.html?module=inline; https://www.nytimes.com/2019/09/24/technology/europe-google-right-to-be-forgotten.html} AI amplifies this pressure. 

Self-regulation will undoubtedly play an important role in the emerging regulatory ecosystem for AI. But self-regulation lacks a critical dimension for effective regulation, namely political accountability.  Self-regulation does not take place in a vacuum of accountability--indeed, many self-regulatory efforts are responsive to politics as they are efforts to ward off direct government regulation. But self-regulation is not formally or transparently subject to accountable oversight, by definition.  

A central goal of our proposal for regulatory markets is to develop a framework of oversight and public accountability for private regulation.  We propose a method for harnessing the power of markets to develop more agile and technically sophisticated forms of regulation that does not require defaulting to self-regulation.  Instead, our approach is to create a new market layer of independent private regulators who are subject to government oversight while simultaneously responsive to the on-the-ground realities of fast-moving, complex, and global AI technologies.

\section{Limits of Existing Regulatory Methods for AI}
Figuring out how to control or channel AI is a technical and system design challenge comparable to the challenge of figuring out how to build and deploy AI in the first place.  The latter challenge is primarily being addressed by markets. Research organizations (for-profit corporations such as Google as well as universities and non-profit organizations such as the Vector Institute), even when they make use of government support, rely heavily on private investors to cover the costs of recruiting top engineering and other talent, buying (and building) powerful computers, conducting experiments, and bringing researchers together for scientific exchange and collaboration. Even China, with massive public sector participation in the economy, will rely heavily on channeling private investment into AI to achieve its national goals of AI dominance \citep{Lee2018}.  

Private investors support the research effort in AI because they anticipate the opportunity to benefit financially (and to some extent personally) from the research. There are (coarse) data that suggest private investors already out-spend governments on certain key areas: consider, for instance, that in 2017 the US National Science Foundation's total spend was about \$6 billion dollars, and that NSF believed this made it responsible for 83\% of the total funding for computer science R\&D that year. By comparison, Alphabet Inc. alone spent \$16.5 billion on R\&D in its 2017 fiscal year.

Current research efforts in the domain of designing regulatory systems, however, take place primarily in the public sector.  The work being done by the National Highway Transportation and Safety Agency to develop a regulatory regime for self-driving vehicles is an example. In this domain, design efforts are driven by the incentives of politics and bureaucracies. This makes them reasonably accountable to their constituents and gives them legitimacy to act on behalf of the public. But it limits their resources to publicly-funded budgets and settings in which the incentives for people to join the project are muted and research expenditures are more limited than they are in the private sector. Moreover, it limits the domain of solutions to conventional methods of regulation: text-based rules, public investigations and monitoring, prosecution of violators in administrative and judicial proceedings, fines and prohibitions. The people participating in the effort are mostly trained in the humanities and social sciences with limited computational expertise or knowledge of technology; few with technical expertise come on board.

Private efforts to develop regulation are playing a growing role in AI. These efforts are coordinated through non-governmental private standard-setting membership organizations (SSOs), such as the ISO and the IEEE, and corporate participation in government standard-setting bodies, such as NIST (which originated as a private organization in 1901 but is now an agency of the U.S. Department of Commerce). All of these organizations, and more, have launched AI initiatives.\footnote{https://standards.ieee.org/news/2017/ieee\_p7004.html; https://www.iso.org/news/ref2336.html} In both public standards agencies and private SSOs, resources from private corporations (regulated entities) are devoted to regulatory development, generally by making participation as a volunteer on standard-setting committees a component of corporate job descriptions.

Even taking into account these private resources, however, there is a tremendous imbalance in the type and volume of R\&D resources between those devoted to regulatory solutions for AI and those devoted to building the technology we need to regulate.

These imbalances are an important source of AI risk.  Consider for example the resources spent by Volvo and other vehicle manufacturers to develop smart systems able to detect when the car was being tested for compliance with emissions standards and adjust the car’s performance to outwit the test.\footnote{https://www.nytimes.com/interactive/2015/business/international/vw-diesel-emissions-scandal-explained.html} In the regulatory race, private investment can generally outpace public expenditure. We should anticipate that AI companies will face tremendous commercial incentives to minimize the impact of regulation on their products and services; only robust and intelligent regulatory regimes will be capable of reigning in those incentives. 

The key to adapting our regulatory systems to keep up with powerful AI is to figure out how to accelerate both the pace and quality of efforts to innovate better ways to regulate.  One way to do this is by making public sector regulation more experimental and responsive, following the lead of industry approaches to developing new goods and services; this is the challenge of ``agile governance'' \citep{WEF2018}.  

Regulatory sandboxes, for example, allow companies or whole industries to develop new goods, services, and processes in a limited and closely-watched domain, without complying with existing regulations, in order both to allow the innovation to develop and to allow regulators to learn about what regulation might be needed before acting. China is characterized by some observers as pursuing this approach to regulation: develop first, learn about the effects, then develop regulation \citep{Lee2018}.  

Policy labs within government engage civil servants directly with the techniques of human-centered design and data analytics to develop new methods of regulation and public service delivery.  

In general, however, increasing the ability to generate innovative regulatory approaches is likely to require moving more of the problem of regulation out of the domain of the public sector and into the domain of markets: creating markets for regulation that attract money and brains (especially engineering brains) to the problem.   

A noted above, to some extent, this is already happening, primarily within private SSOs. But these membership organizations tend to behave more like political bodies than the private sector \citep{Birdthistle2013}. They operate on the basis of committees composed of members, voting, and consensus. The analogy to political bodies is not perfect: many do sell their standards and certification services and so there is some scope within which we can think of these organizations as competing. But the nature of this competition is that the standards bodies compete to be adopted voluntarily by a large enough segment of a relevant industry that they become “the” de facto global standard. Thus they are not competitive markets in the full sense. Moreover, once the competition is resolved in a given domain, there is the potential for stagnation. Continuing competition is key to driving innovation.

As important as it is to recruit private sector incentives to the problem of producing more effective forms of regulation to keep up with technology and globalization, however, it is also critical to ensure that regulation is legitimately anchored in the interests of relevant communities of people. Current private standard setting is responsive to the public interest only indirectly--as channeled through the interests of corporations in maintaining a good reputation for corporate social responsibility with their customers, employees, and local governments. Moreover, participation in private standard setting is voluntary. But private regulation of AI should not be controlled by corporations that profit from AI or a few powerful people. Nor should it be entirely voluntary. It should have mandatory components and be subject to more direct oversight than is true of private standard setting today. Accomplishing that objective requires structuring these markets for regulation in a way that is accountable to the public sector.  

The other shortcoming of public sector efforts alone to address the regulatory gap for AI is the need for truly global--transnational--solutions. Public sector regulation is currently organized primarily on a nation-state basis.  International agreements between states can coordinate regulation across countries, but the implementation of global standards still happens through domestic regulatory regimes. This is another reason that private SSOs have emerged as such an important player in technology regulation: they are capable of generating standards that are developed at a transnational level and implemented as such across multiple jurisdictions. The ISO, for example, is a membership organization that is composed of country SSOs, one for each member country.

\section{Global Regulatory Markets}
Increasing the ability to generate innovative regulatory approaches is likely to require moving more of the problem of regulation out of the domain of the public sector and into the domain of markets: creating markets for regulation that attract investments of human and financial capital in regulatory innovation. We propose the following model of regulatory markets, building on new governance models that incorporate non-governmental regulators and changing roles for private entities. 

There are three principal actors in this model: The targets of regulation, private regulators, and governments. Targets are businesses and other organizations that governments seek to regulate. In the AI context, these are the companies or organizations building and deploying AI products or services. Private regulators are for-profit and non-profit organizations that develop and supply regulatory services which they compete to sell to targets. Governments require targets to purchase regulatory services (entering into a regulatory contract with a private regulator) and directly regulate the market for regulatory services, ensuring it operates in the public interest. 

\begin{figure}
\centering
\includegraphics[width=\textwidth]{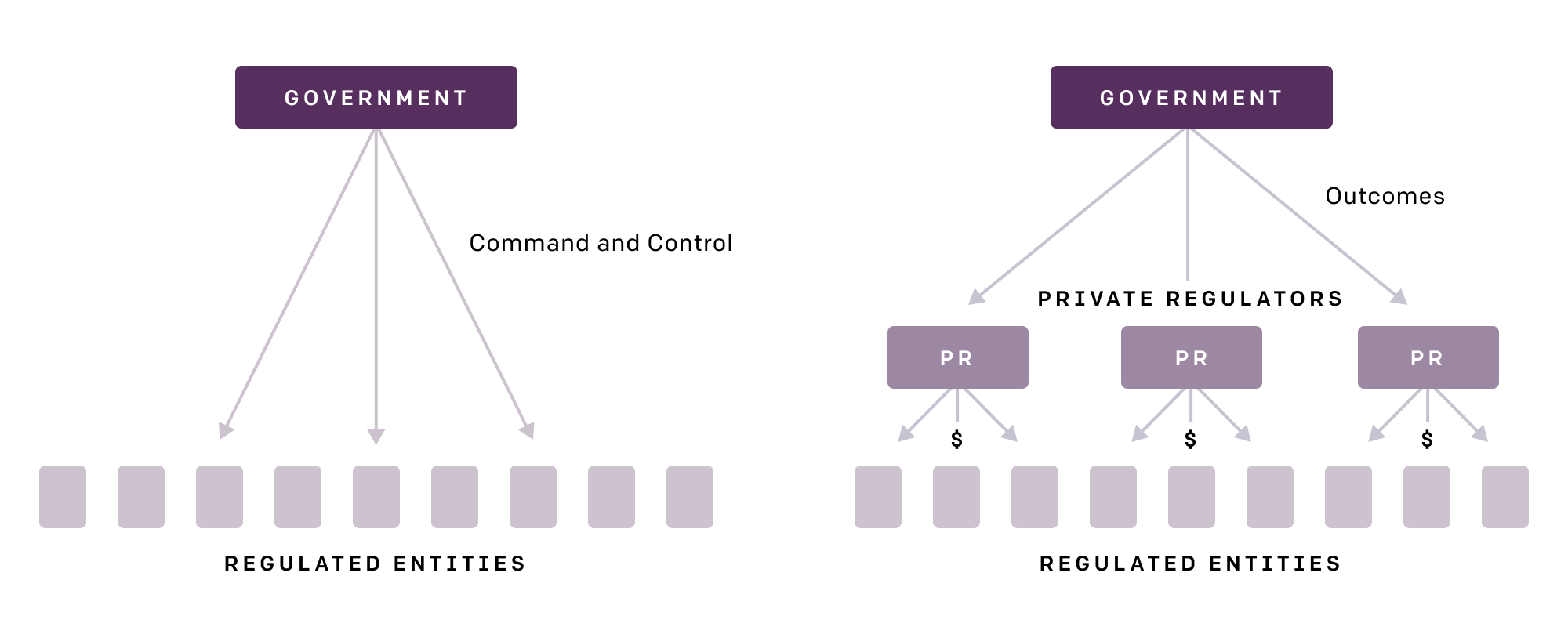}
\label{fig:Figure 1}
\caption{In conventional regulation, shown on the left, the government directly regulates entities. With regulatory markets, shown on the right, a private regulator directly regulates the targets that have purchased its regulatory services, subject to oversight by government to ensure regulators are achieving outcomes set by government. It does this by developing regulatory procedures, requirements, and technology.}
\end{figure}

Private regulators could employ, but would not be limited to, conventional means of regulation; the private regulator might also develop technologies that directly control or shape the business decisions of the targets it regulates. They would gain their authority to regulate via the regulatory contract with the target and authorization from governments to collect fines or impose requirements on the targets that submit to their regulatory system.

Here are a few examples: 
\begin{itemize}
\item A private regulator of self-driving cars might require self-driving car companies to allow the regulator access to data produced by the vehicles and then use machine learning (ML) to detect behaviors that raise the risk of accidents beyond thresholds set by the regulator. The private regulator might bring these to the attention of the target and require risk-assessment; or it might develop technology that allows the regulator to modify the algorithms or data sources used by the target’s vehicles. 

\item A private regulator in the banking industry might require a bank using ML to analyze customer data and develop new products to implement differential privacy techniques \citep{Dwork2014} to minimize the likelihood that a customer is harmed by the use of their data. The regulator could prescribe the specific techniques/algorithms to use; or it could establish a procedure for the banks that it regulates to propose techniques that survive tests conducted by the regulator.

\item A private regulator of developers of drones equipped with facial recognition systems might require companies to implement particular cybersecurity features to ensure their models are not discoverable by malicious users. The regulator might also create systems that enable people to raise flags about drone behavior to detect malicious use.
\end{itemize}

Regulatory techniques developed by private regulators might include hardware, risk assessment tools, information processing systems, conflict or complaint management procedures, and so on.

In order to participate in the market by selling regulatory services to targets, private regulators must be first licensed by the government in the jurisdictions in which they wish to operate. In any given domain, multiple regulators are licensed so that they compete to provide regulatory services to targets. Targets must choose a regulator but they have the capacity to choose, and switch, regulators. They do so by comparing across regulators in terms of the cost and efficiency of the services provided by regulators.

Private regulators do not compete, however, on the quality of their regulatory services, that is, the extent to which they achieve public goals. This is because in order to obtain and maintain a license, regulators must demonstrate their regulatory approach achieves outcomes that are mandated by government. Outcomes are metrics or principles set through the bureaucratic processes of the public sector. They are the mechanism by which the delegation of regulatory oversight of target to private actors is made legitimate.

For example:
\begin{itemize}
\item In the self-driving car context, governments could set thresholds for accident rates or traffic congestion. They could establish principles such as maintaining public confidence in road safety. 
\item In the banking industry, governments could set thresholds for access to credit by consumers. They could establish principles such as traceability of transactions and maintenance of confidence in the stability of financial markets. 
\item In the context of facial recognition use in drones, governments might establish thresholds for the likelihood that software could be accessed by malicious users. They could establish principles such as realistic consumer consent to recognition.
\end{itemize}
The key here is a shift by government to establishing the goals of regulation, rather than the methods of achieving those goals. Methods are developed by the private regulators, and then tested by governments. This testing would occur through a combination of upfront evaluation of the capacity for a regulator's system to satisfy government goals and ongoing auditing and oversight: measurement of outcome metrics and assessment of the achievement of principles. For example, in the self-driving car setting, governments may develop techniques to track accident and congestion rates and assess the contribution of a particular regulator to excessive accidents or congestion. In banking, governments could conduct periodic audits of random samples of transactions from the targets of a particular regulator to determine the incidence of money-laundering. In drones, governments might stress test a regulator's procedures by employing adversarial efforts to infiltrate algorithms or data. 

Regulators that fail to pass the tests set by governments would risk having their licenses suspended, conditioned, or revoked. This requires governments to regulate to ensure that the market for private regulators is competitive, ensuring that there is sufficient scale in a given domain to support multiple regulators (possibly restricting the share of the target market that a given regulator can service) and that targets have the capacity to switch regulators with relative ease. This obtains the benefits of competition between regulators, spurring them to invest in developing more effective and less costly means of achieving regulatory objectives. 

Protecting the integrity of the market requires  that private regulators are independent entities. They must be neither formally nor informally controlled or captured by targets. This is critical to ensure that the regulators’ incentives are to produce excellent regulation, not collaborate with targets to reduce the quality of regulation.

Private regulators, ideally, are licensed by multiple governments, each implementing their own outcome requirements. “Search Oversight Inc.”, for example, could be licensed to regulate the relevance of and consumer control over search engine results in many of the countries in which Google does business. Similarly, Microsoft's Bing might be regulated by a single regulator in many or most of the countries in which it operates. But Google might be regulated by a different regulator than Bing. Both, however, would be held by their regulators to achieving the same ultimate regulatory outcomes. 

\section{Locating Regulatory Markets on the Landscape of Regulation}

Our proposal extends these existing private regulatory models in three ways. First, it envisions privately-devised standards being developed subject to government oversight. IEEE, for example, would not have the final say on what the standards should be; it would need to design those standards with a view to meeting government-set (politically accountable and legitimate) outcomes and principles. Second, the model envisions a shift from voluntary to mandatory adoption of standards. Companies would have the option of choosing whether to adopt IEEE standards or some competing regulatory regime in some domain, for example, but they would not have the option of remaining unregulated. Third, private regulators would have access to a wider range of enforcement tools than they do now. The responses to the Bangladeshi work safety disasters, for example, depend on the use of contract tools to enforce compliance. Under our model, regulators would have broader powers such as to the capacity to impose fines, require audits and supervision, and revoke licenses to operate.

Critically, regulatory markets are not a means for self-regulation or delegation of regulatory oversight to a (target) industry. Targets are not regulating themselves; they are submitting to the regulatory regime of an independent private regulator. Instead of Google developing and implementing internal procedures to protect a government-established right to be forgotten in search results, for example, Google would submit to the procedures (or implement the technology) developed by a third-party company or organization (“Search Oversight Inc.”). 

\section{Benefits of regulatory markets}
The reason to build regulatory markets is to harness market incentives to invest in the development of new regulatory technologies. Powerful incentives are driving corporations and nations to invest in building and deploying AI as fast and as broadly as possible. We will require equally powerful regulatory systems to ensure that AI systems remain safe and beneficial. We will almost certainly, for example, need AI to assist in monitoring and regulating AI systems. The only way to get robust and durable investments in building regulatory AI is by creating a market incentive: rewarding such investments with the achievement of profit and personal mission. 

A key dynamic that we hope to capture is the start-up dynamic by which expertise in a domain migrates out of large organizations into new entrepreneurial ventures. Much of the expertise about the nature of the challenges we face with AI resides inside AI research labs--almost all of which are housed inside corporations. Many of these corporations are also financing research on how to build safe and responsible AI. By creating a market outlet for such expertise, we create the potential for some of this expertise to migrate out of the corporations and into concrete efforts to build industry-wide regulatory tools. In this vision, the AI safety/policy team at OpenAI, for example, could become a start-up in the regulatory market: transforming its research into a concrete set of regulatory tools and services that it first proves (to governments) achieve specific AI regulatory goals and then sells to AI companies required by governments to submit to regulatory oversight.

Another benefit of the regulatory markets model is that it fosters the development of a robust and independent regulatory sector. This can help balance the growing power of the large tech companies that are at the forefront of AI development. It creates a middle option between the unsatisfactory choice between government-led regulation on the one hand and tech company self-regulation on the other. 

Private regulators may be able to govern data more effectively. Currently, a critical constraint on government regulation is that sharing data with governments creates the real and perceived risk of the misuse of data by governments. Commercial entities, on the other hand, are able to engage in more robust data-sharing arrangements, governed by contract and intellectual property rights. This would allow private regulators to engage in more direct monitoring of data held within target companies, and creates the possibility for more creative technological integration between regulator and target. 

\section{Risks and limitations of regulatory markets}

Regulatory markets come with risks around the challenge of ensuring that private regulators are competitive and independent of the entities they regulate. 

Competition might fail because there is insufficient scale to support multiple regulators--if there are only two or three companies involved in developing a particular type of AI, it will be difficult to sustain a competitive market of regulators, each of which needs to regulate multiple entities and each of which needs to be at risk of losing market share in order to prompt continual investment in better regulatory technology. Even where there is sufficient scale, competition might not emerge if a single regulator gains too much market share or if the costs of switching regulators (the threat that keeps competitive pressure on regulators) is too high. Competition might also fail if regulators collude. Some of these concerns can be addressed through the design of the regulatory environment imposed by governments: antitrust and competition law could protect against the monopolization of the regulatory market, but robust competition might require additional protections  such as limitations on market share or rules (about data portability and sharing, for example) to reduce switching costs. 

The independence of regulators will require close attention. Regulatory capture is a known risk in existing government-led regulation--both through explicit corruption and through more subtle mechanisms: campaign finance and lobbying, overlapping industry and regulator professional networks, the dependence of regulators on information supplied by industry, etc. \citep{DalBo2006} Regulatory markets put an additional layer between governments and industry. This creates a risk that private regulators, which are trying to sell their services to AI companies, will collaborate with those companies to cheat on government goals. Protecting the integrity of regulation will require governments to monitor the results achieved by private regulators and to have effective threats to condition, suspend, or revoke the licenses of regulators that skimp on performance in order to win the business of targets. This is a transformation of the existing problem of regulation: regulation is only as good as the capacity and willingness of governments to regulate. With good design it is possible that regulatory markets make it easier for governments to regulate: multiple regulators means multiple sources of data and industry expertise. Participants in the regulatory market will also have an incentive to monitor the performance of their competitors, perhaps exposing cases in which competitors are ``cheating'' on regulatory outcomes to achieve markets share. At the same time, instead of regulating, for example, 1000 companies engaged in the production of AI systems in a given domain, government will be focused on regulating perhaps 5 or 10 regulators. 

Another risk is that governments will come under political pressure to displace private regulators in response to high-profile accidents or crises. Legislators might then encroach on the domain of the private regulator--moving away from specifying and evaluating outcome metrics and principles to dictating more of the details of regulation. If this is anticipated it could undermine confidence in the power of the private regulator and diminish the willingness of targets to cooperate with them. 

Finally, although an important benefit of multi-jurisdictional regulators is that less-wealthy jurisdictions can free-ride on the regulatory oversight of wealthier ones, a general free-riding problem could result in governments as a whole under-investing in oversight.

\section{Comparison to Existing Hybrid Regulatory Models}

We can learn from the successes and failures in existing uses of private actors in regulation to understand more about the possibilities and pitfalls of regulatory markets for AI and how best to design those markets where feasible. In this section, we review two particularly powerful examples where hybrid public-private governance models have failed and two examples that shows promise as models. 

\subsection{Credit Rating Agencies and the Financial Crisis}
In the U.S. private credit rating agencies (CRAs)--Moody's, Standard \& Poor, Fitch, and a handful of smaller for-profit companies--supply ratings of the bonds issued by governments, for-profit companies, and non-profit organizations.   Private credit rating agencies emerged in the early 1900s as a market response to investor demand for information on the credit risk associated with issuers. Today, all CRAs use some form of the AAA to C- scale introduced in 1909 by John Moody to rate railroad bonds. Investors originally paid for these ratings and there was no involvement from government.  During the 1930’s, however, in response to financial crises, governments began to use ratings to regulate banks and other financial institutions.  They did so by, for example, prohibiting banks from purchasing securities rated as 'speculative' (below BBB) by the private credit rating agencies.  Then, in 1973, the SEC issued a rule linking the capital requirement imposed on broker-dealers (the amount a broker-dealer must maintain in liquid assets) to credit ratings assigned by what the SEC deemed to be a Nationally Recognized Statistical Rating Organization (NRSRO)--a group that consisted at the time of Moody’s, S\&P, Fitch, and Duff \& Phelps.  Following this rule, references to credit ratings supplied by this group began to appear in a wide variety of financial regulations: banking, securities, insurance, pensions, and real estate.  Over time, the business model of the credit rating agencies shifted. Initially, agencies were paid by investors to supply information. In the 1970s, however, agencies switched to an issuer-pay model, in which issuers paid on fees for the rating they needed to sell securities. \citet{Partnoy2009} calls the authorization issuers bought from agencies a “regulatory license.”  Only with the required credit rating could these issuers sell their debt-based instruments to regulated entities such as banks, mutual funds, or pension funds.  

The quality of credit ratings has been in doubt since at least the late 1990s \citep{Partnoy1999}. Orange County was granted the highest-possible rating by S\&P and Moody’s for a bond issue just a few months before filing bankruptcy in 1994. Enron was similarly highly-rated right up to its bankruptcy in 2001. The failure of credit ratings to accurately reflect risk, however, was nowhere more spectacularly on display than in the financial collapse of 2008, in which rampant gaming of credit ratings in asset-backed securities resulted in the sub-prime mortgage crisis and a rapid downgrading of entire classes of securities from AAA to junk status practically overnight.  A Congressional inquiry pinned the blame for the collapse squarely on the credit rating agencies \citep{Commission2011}. 

The failure of credit rating agencies in the sub-prime mortgage crisis is a cautionary tale for regulatory markets.  The lessons rest in an appreciation of the regulatory oversight of the credit rating agencies, or more precisely, the lack thereof. When the SEC first designated “approved” credit rating agencies as NRSROs in 1973, it did so without the creation of any direct regulatory oversight of these agencies. Indeed, the SEC \emph{exempted} NRSROs from liability for the accuracy of their ratings by shielding them from lawsuits brought by those who relied on them. The credit rating agencies have successfully characterized their ratings as mere ``opinions'' about creditworthiness. Courts have supported this lack of accountability by accepting arguments from credit rating agencies that they are protected against liability for their ratings by principles of free speech. Even after the clear weaknesses in credit ratings became politically salient in the early 2000s, the 2006 Credit Rating Agency Reform Act, while giving the SEC greater oversight powers, preserved the historical exemption from liability and prohibited the SEC from regulating ``the substance of credit ratings or the procedures and methodologies by which any [NRSRO] determines credit ratings'' \citep{Partnoy2009}.\footnote{The 2006 Act introduced a registration scheme for NRSROs and required registered NRSROs to provide the SEC with regular information and certifications from users of its ratings.  The SEC was authorized to suspend a registration for failures to provide the required information or if the SEC made a determination that an NRSRO ``fails to maintain adequate financial and material resources to consistently produce ratings with integrity.'' 15 USC 78o-6 15E(d)(A)(5). The SEC was expressly prohibited, however, from ``regulat[ing] the substance of credit ratings or the procedures and methodologies by which any nationally recognized statistical rating organization determines ratings.'' 15 USC 78o-6 15E (c)(2).}  After the 2008 crisis, Congress ostensibly tightened regulatory oversight in the 2010 Dodd-Frank Act, creating the Office of Credit Ratings (OCR) within the SEC. The OCR was tasked with reviewing each of the (now 10) NRSROs at least annually and publishing an annual report. The report, however, only provides summary information and does not disclose the identity of a credit rating agency found to have violated regulatory requirements \citep{Partnoy2017}. Moreover, the office does not have any independent authority to bring enforcement actions and an overwhelmed SEC has engaged in little enforcement activity. Finally, although Dodd-Frank removed the protection credit rating agencies historically have enjoyed from lawsuits based on faulty ratings, the SEC has effectively reintroduced such protections \citep{Partnoy2017}).  

A core lesson is clear:  regulatory markets cannot operate effectively if they are not effectively overseen by government agencies with budget and the capacity to resist capture. Without strong oversight, private regulators may only `compete' by lowering their standards. (Evidence of this is provided by the example of S\&P, which recovered from a negative shock to its reputation--caused by errors requiring withdrawals of ratings--by issuing more optimistic ratings than its competitors \citep{Baghai2019}.) Our proposal rests on the design of effective outcomes-based regulation of private regulators.  

A second lesson from the experience with credit rating agencies is related to the capture risk. The market for approved credit rating agencies is not competitive. Although there are now 10 approved NRSROs, the market is heavily dominated by just two: Moody`s and S\&P. Another key feature of the design of regulatory markets needs to be active efforts to protect competition, limiting market share if needed.  

A third lesson can be taken from various diagnoses of why efforts to reign in the credit rating agencies have failed. As several scholars have noted, it has proved tremendously difficult to reduce the reliance on NRSRO ratings because of how deeply integrated they have become in so many regulatory and financial schemes.  This undermines the credibility of any threat to deprive a CRA of approved status. The lesson for the design of regulatory markets for AI is that close attention must be paid to the capacity to act on a threat to penalize a private regulator that is found not to be producing the results required by government. 

\subsection{Self-Regulation and the Boeing 737 MAX Crashes}
In the United States, the Federal Aviation Administration is tasked with overseeing civil aviation. The FAA is part of the Department of Transportation, which has the goal of ensuring ``a fast, safe, efficient, accessible, and convenient transportation system.'' As part of that remit, the FAA oversees airlines. 

In 2018 and 2019 two Boeing ``737 MAX'' airlines crashed as a consequence of regulatory failure. Specifically, software systems introduced by the plane manufacturer Boeing led to poorly documented behaviors by the plane in rare cases, which--combined with improperly trained pilots--caused crashes to occur. 

The suspected root of this failure was a lack of oversight by the FAA into the development of the maneuvering characteristics augmentation system (MCAS) within the planes. This was a consequence of the FAA delegating a large portion of regulatory oversight to aviation companies such as Boeing themselves, through the ``Organization Designation Authorization'' (ODA) program.\footnote{\url{https://www.faa.gov/news/testimony/news\_story.cfm?newsId=23514&omniRss=testimonyAoc&cid=105\_Testimony}} 

The ODA provided a formal mechanism for the FAA to delegate certain oversight activities to organizations and  was created in response to activity in the aviation sector outpacing the FAA’s own ability to effectively regulate the sector. Prior to 2004, the FAA appointed Designated Engineering Representatives (DERs) to perform oversight of a given product within an aviation company, and this person--though being paid by the aviation company--would report directly to the FAA. Under the ODA change, the FAA instead recognizes an ODA organization within an aviation company, and this ODA selects staff who are themselves managed by the company. The FAA ostensibly oversees the ODA via spot checks, but it engages in little direct oversight of the personnel within the ODA. 

Both the DER and ODA system have had problems: A 2011 investigation conducted by the Department of Transportation's Office of the Inspector General (OIG), prompted by a request from a congressman concerned about lack of oversight, found numerous instances in which manufacturers had appointed or retained DERs with poor performance history or over the objections of FAA engineers. The report also highlighted that some oversight officers did not even track DER personnel by name and hence could not identify poor performers who lacked either technical skill or appeared to be acting in the manufacturer's interest at the expense of compliance. An analysis of the FAA's audits for 2005-2008 found 45 instances in which the FAA had not caught failures in certifications of safety systems, such as a complete absence of  ``evidence that critical tests on a new aircraft engine component were ever performed'' \citep{Inspector2011}. A subsequent report in 2015 found that ``one inspector responsible for oversight of nearly 400 manufacturing personnel performing work on FAA's behalf reviewed the work of only 9 personnel during fiscal year 2014'' \citep{Inspector2015}. 

Boeing’s implementation of the ODA caused it to change how it managed employees within its safety organizations. Media reporting following the 2018 and 2019 crashes indicates that though numerous people within Boeing had identified safety issues with regard to the software systems on the 737-MAX, they had been discouraged from reporting these issues by their (Boeing) managers. Additionally, people that worked for the FAA were being pressured by their own FAA managers to rapidly qualify aspects of the plane for safety, despite lacking both the staff and time to do a good job \footnote{https://www.seattletimes.com/business/boeing-aerospace/failed-certification-faa-missed-safety-issues-in-the-737-max-system-implicated-in-the-lion-air-crash/}. These reports are consistent with a 2012 OIG investigation that substantiated allegations from FAA staff responsible for overseeing Boeing that FAA managers were ignoring or overriding efforts to hold Boeing accountable. Staff recommendations to remove a Boeing ODA administrator and address conflicts of interest, for example, were overturned, moves seen by staff as ``evidence of [FAA] management having too close a relationship with Boeing officials'' \citep{Inspector2012}.

It has been clear since 2011 and was put in stark terms by the 2015 OIG investigation that there has been little meaningful oversight of Boeing for many years \citep{Inspector2015}. The 2015 inquiry found that oversight was not risk-based, focused on meeting minimal checklist requirements and minor paperwork errors instead of safety-critical systems, and lacked an appropriate staffing model to ensure adequate resources for inspectors. Indeed, Boeing, with a dedicated FAA oversight office for its ODA, was not in the staffing model at all; the determination of how many inspectors to allocate was not based on data or risk analysis, leaving FAA with no ability to assess the adequacy of staffing in that office. Gaps were especially large in the supply chain, with ``FAA performing oversight of only 4 percent of personnel conducting certification work on FAA's behalf in fiscal year 2014'' \citep{Inspector2015}.

The lessons from the Boeing 737 MAX disasters are still being learned but we can see at least two so far for regulatory market design. One is that the pressure to devolve regulatory duties to private actors is intense and has produced extensive reliance on self-regulation--even in safety-critical contexts such as aviation. Our proposal for regulatory markets can improve on this by shifting those duties to independent regulators, rather than the targets of regulation themselves. Second, however, any regulatory scheme requires adequate oversight by governments. The insufficiency of FAA oversight practices is a cautionary tale for our regulatory markets proposal, emphasizing the need for a sustainable funding model that accords with the true cost of regulation. Under our proposal, at least some of this cost would be priced in the market as the cost of regulatory services, rather than being entirely dependent on the politics of taxation and government budgets. 

If, instead of delegating regulatory duties to Boeing (or some other airplane manufacturer or engineering entity), one (or several) of those manufacturers had spun-out a safety assurance startup to operate as an independent private regulator, we could imagine this startup being overseen by the FAA with the goal, for instance, ``to guarantee the safety of new flight platforms, and guarantee that typical pilots can be trained to use the platform within a day of study.'' The startup, given that objective, would be incentivized to not only analyze the MCAS system for particular failure modes and safety issues, but also to validate that human pilots could be trained against it. This startup would face significant economic pressure to develop effective methods for overseeing such systems, but it wouldn’t face the same kind of conflict of interest that airline employees face when having objectives (for instance, safety) that conflict with direct orders from their managers (qualify this plane quickly). 
\subsection{Medical Device Safety and the Pressure to Harmonize Regulation}
We now turn to some more promising examples of hybrid public/private regulation, close to our regulatory markets proposal, that have recently emerged. The first is in the regulatory arena for medical devices. This example sheds light in particular on the pressure to build regulatory schemes that can regulate at global scale.

Medical devices range from the simple (tongue depressors) to the safety-critical (cloud-connected pacemakers). Regulators in most countries have detailed schemes, for devices deemed more than minimally risky, that govern all phases of development, production, and marketing. In response to the needs of companies attempting to sell their devices internationally, facing complex and conflicting regulatory requirements, in 1992 a consortium of regulators from the U.S., Canada, Australia, the E.U., and Japan formed the Global Harmonization Task Force (GHTF) consisting of regulators and industry representatives with a goal of developing a uniform regulatory model for adoption in member countries. Harmonization across jurisdictions has long been a goal of global efforts to reduce regulatory burdens on industry. The task force generated a regulatory model but acknowledged in 2011 that, 18 years after the project began, the model had not succeeded in achieving uniform regulatory practices in member countries. In addition, the group faced the challenge that over time its membership was too narrow: it failed to reflect changes in the global market, excluding, for example, Asian countries, and it excluded stakeholders other than manufacturers, such as healthcare providers, academics, and consumers. In 2011, the GHTF disbanded.  

The GHTF was replaced in 2011 by a new consortium of regulators known as the International Medical Device Regulators Forum (IMDRF), consisting of regulators from the original member countries of the GHTF plus, eventually, regulators from Brazil, Russia, China, Singapore, and South Korea and observers from the World Health Organization. Learning the lesson of the failed GHTF effort, instead of aiming at harmonized legislation to enact convergent regulatory systems in member countries, the IMDRF has focused instead on technical convergence (such as a uniform device identification system and standards for cybersecurity) and regulatory processes that don't require legislative change. 

A major initiative of IMDRF has been the introduction of the Medical Device Single Audit Program (MDSAP), which was piloted in 2014-2015 and became fully operational in 2017. The goal of this program is to create a scheme in which a medical device can be audited by a single organization for compliance with the (quality management) standards of any of the countries in which it will be sold.  Countries participating in the program agree to accept the audit report of the single auditor as meeting the certification requirements of their regulatory scheme.  

MDSAP authorizes private auditing organizations (AOs) to audit medical device manufacturers according to protocols established by MDSAP. MDSAP then audits the auditors (engaging in what are called "witnessed audits" with reporting back to MDSAP and the AO) to ensure that they are completing audits as required. The quality management standards reflect those implemented in all jurisdictions in which the manufacturer seeks to distribute; in many cases, jurisdictions have converged on a quality management standard promulgated by the International Standards Organization (ISO 13485). Device manufacturers purchase the auditing services of an AO and participating regulators (currently Australia, Brazil, Canada, Japan and the U.S.) accept this single audit report as satisfying their requirements. As of August 2018, there were 14 AOs either fully recognized or in the probationary period leading to full recognition.  Participation by device manufacturers grew more than ten-fold from 222 at the start of the operational phase in January 2017 to 2,711 by August 2018.  

This model tracks many features of our regulatory markets proposal. There is a global competitive market for auditors, with regulated entities paying a competitive price for services. These revenues are used to cover the cost of audits.\footnote{We have not yet been able to confirm the funding model for the MDSAP oversight body itself, whether it is funded by auditing organizations--thus priced into the fees they charge in the market for their services--or by contributions from participating governments.} The government bodies focus on regulating the regulators--auditing the auditors.  The model is not a full representation of our proposal, in that the standards--which are determined by the government regulators--are not necessarily outcomes-based; they may be highly prescriptive. But the system enables a global market for regulation in which individual countries are not obliged to adopt the same regulatory standards. They may have incentive to do so, in order to reap the benefits of a more efficient auditing market that has developed to audit for a particular set of standards, and indeed a few of the participating countries have adopted the privately-devised quality standards developed and sold by the International Standards Organization. Even with divergent standards, however, manufacturers can work with a single entity (auditor) to achieve regulatory compliance that satisfies the requirements of multiple jurisdictions. 

It is too early to assess the efficacy of the MDSAP.  But with the credit rating agencies and 737MAX failures in mind, we can see the potential for greater success.  Unlike the credit rating agencies, there is a formal oversight mechanism to discipline private regulators. And unlike the FAA weaknesses that spelled disaster with the 737MAX, there is a third-party independent regulator and there are formal, publicized procedures for regulating the regulator. Moreover, because the scheme has been set up as a global consortium, there is less risk of capture: even if a regulator (auditor) can capture one government, it is unlikely to be successful at capturing five or six or more.  The formal structure of the market also helps here: the audits are provided by entities that have a global market opportunity, which both encourages investment in discovering methods of accomplishing audits in a cost-effective way and creates a greater penalty for a loss of reputation or loss of formal authorization to perform audits. Last, by pricing at least some of the cost of regulation in the market for regulatory services, the system is better protected against political pressures on the budgets of government regulators.   

The MDSAP model shows the feasibility of building global markets for independent regulatory services that avoid the trap of harmonization efforts--which sank GHTF--and reduce the risks of capture and failed regulation.  All while recruiting the market to provide more nimble responses to complexity and innovation.

\subsection{Legal Markets in the United Kingdom and Legacy Constraints on Competition}
As a final promising example, we turn to what may seem to be an unusual case: markets for legal services.  Historically, markets for legal services have been self-regulated by providers--lawyers--acting through voluntary organizations such as bar associations and law societies. In 2007, however, the U.K. adopted a novel scheme, prompted by concerns about capture and a lack of competition, which tracks all of the features of our regulatory market proposal.  

The 2007 Legal Services Act (LSA) created an independent government-appointed body, the Legal Services Board (LSB)to govern the provision of legal services in England and Wales.  Instead of regulating lawyers, however, the LSB was authorized only to approve entities that applied to it for authorization to regulate lawyers. The LSA set out governing principles--regulatory objectives--to guide the LSB's determination of whether to approve a regulator. The LSB in turn is required to design and implement an adequate scheme for overseeing the regulators. Regulators report to the LSB and the LSB retains the right to intervene in the exercise of regulation by a private regulator, or seek to cancel the authorization to regulate, if the regulator appears likely to fail to achieve the regulatory objectives.  

Under the 2007 scheme, the initial regulators included legacy regulators who had previously operated as self-governing bodies, but only after those bodies had separated out their advocacy and regulatory functions. The Law Society of England and Wales, for example, which had been the self-governing body for solicitors, spun out a separate entity called the Solicitors Regulation Authority, which became an approved regulator under the LSA. Currently 9 regulators are approved to regulate individual providers; some of these regulators are also approved as licensing authorities capable of licensing entities (known as ``alternative business structures'' which permit solicitors and barristers to be employed by or contract to provide services with people and entities other than solicitors and barristers.\footnote{These are business models that were prohibited by the self-governing regimes that predated the LSA, and which continue to be prohibited in most jurisdictions in the world, including the U.S. and Canada. \citep{Hadfield2020}} 

Any person or entity that wishes to provide one of six activities designated as ``reserved'' under the LSA--including representation in higher courts or filing documents in court, for example\footnote{This designation of reserved activities is much narrower than it is in the U.S. and Canada, where all work requiring the application of legal knowledge to individual circumstances must be done by a licensed lawyer \citep{Hadfield2020}.  Unreserved--unregulated--activities in England and Wales include all forms of legal advice, the drafting of legal documents such as contracts and wills, and some representation in lower courts.}--must obtain a license from an approved regulator and comply with the regulatory scheme its regulator has devised. Fees for these regulatory services are paid by the licensed individual or entity to the regulator; the LSB retains authority to regulate these fees. 

A primary motivation for the development of these regulatory scheme in the U.K. was to address the problem of regulatory capture. A 2001 review of the self-regulating professions (primarily solicitors and barristers) by the U.K. Office of Fair Trading identified several rules imposed on legal practice that have an adverse effect on competition and consumer welfare:, attributable to the fact that ``the professions are run by producers largely on behalf of producers''\citep{OFT2001}. After the professions failed to make any adjustments to improve competition, a follow-on report authored by a banker, recommended the regulatory regime ultimately enacted in the LSA \citep{Clementi2004}. A core concern of this report was to increase competition between providers--today, three professions are all authorized to provide any of the six reserved activities, and other groups are authorized to provide a subset so that there is broad scope for competition between different types of professionals (barristers, solicitors, legal executives, accountants, and so on.) In this sense, the LSA responded to the problem of regulatory capture with the technique we are advocating here: the creation of competing regulators, overseen by a government body responsible for ensuring that the regulatory market was harnessed to regulatory objectives set by Parliament (including the objective of a competitive market for legal services.)

New business models and providers in the U.K. legal services market relatively quickly gained significant market share--about a third within three to six years of the new regulatory scheme coming into effect. New types of providers have shown greater propensity for innovation and achieve higher rates of customer satisfaction \citep{LSB2015}. But the levels of innovation and investment that were contemplated have not yet arisen, and dissatisfaction with the regulatory regime continues. A key challenge appears to be that the regime was built on legacy self-governing regulators and despite formal independence, these regulators have continued to regulate in ways that are largely a continuation of the methods used historically. That is, there has been little regulatory innovation. 

The reason for low innovation appears to be that competition between approved regulators, while available in theory, is very limited in practice. The legacy regulators regulate based on a professional title: solicitor, barrister, legal executive, chartered accountant, etc. Access to the title is based heavily on highly prescribed education or training and apprenticeship requirements. These requirements make switching between regulators prohibitive for individual licensees, blunting any competitive threat. Entities face a greater opportunity for switching regulators, but for several reasons this has not yet produced much regulatory competition: entities only gained access to the market in 2012, multiple licensing authorities only came on line after 2017, and entities are still only about 10\% of the market. 

The lesson for other efforts to implement regulatory markets is that achieving regulatory competition requires close attention to the design of the market. The benefits of regulatory markets are only achieved if there is true competitive pressure to improve on regulatory techniques.  The market of a given private regulator needs, therefore, to be contestable.  This requires relatively low switching costs. The U.K. system ended up with multiple approved regulators not because of an intention to create a regulatory market; this was instead a consequence of the legacy of multiple legal professions that had self-regulated. Had the regulatory design paid more attention to the goal of creating competitive regulators, the problem of switching costs would likely have been recognized and relatively easily addressed.  

\section{Prototype: Verifying Adversarial Robustness}

To demonstrate how regulatory markets might work in practice, we describe in this section a prototype implementation to address a known risk in the AI domain: adversarial attacks on AI models. 

\begin{figure}[!h]
    \center{\includegraphics[width=\textwidth]
    {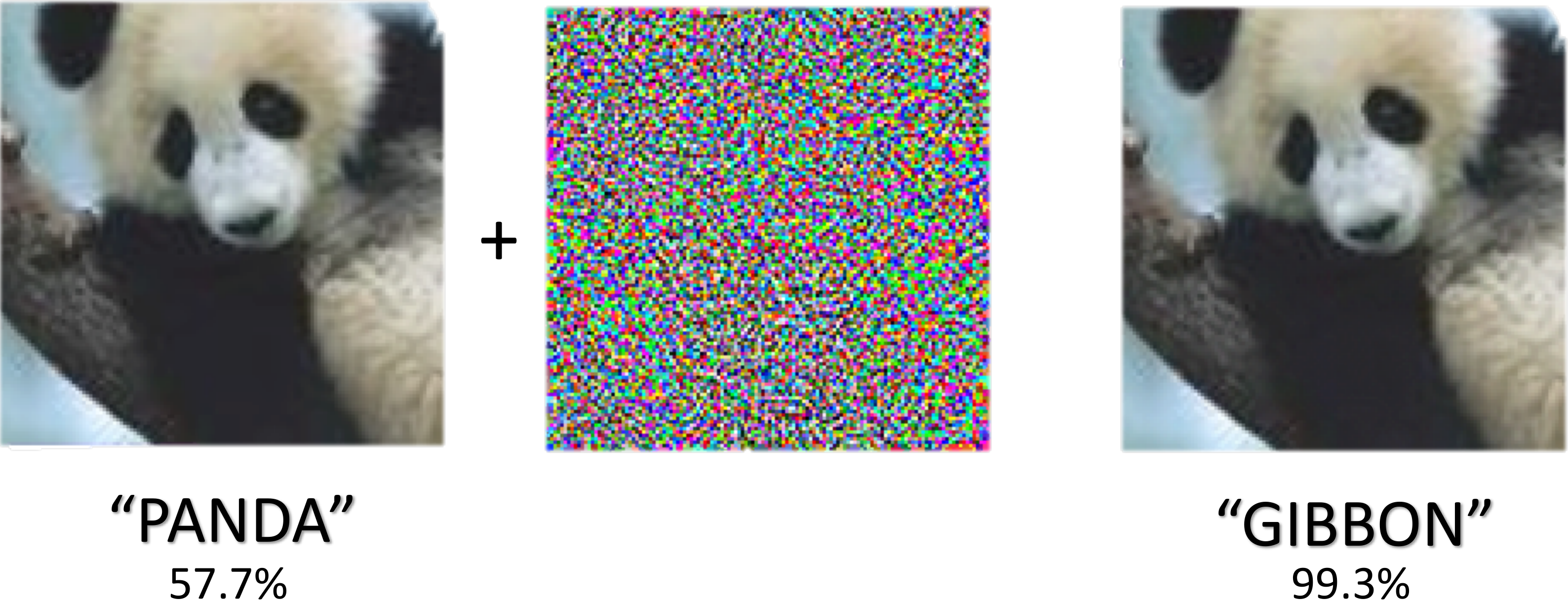}
    \caption{\label{fig:Figure 2} In this example of adversarial attacks in image classification from \protect\citet{goodfellow2015} the model classifies the image on the left as a ``panda'' with reasonable confidence. But if presented with an image that is constructed from the initial image plus a tiny amount of (appropriately chosen) perturbation, it classifies the image as a ``gibbon'' with near certainty.}}
   \end{figure}

With normal training, well-performing deep learning models can easily be fooled by specifically (``adversarially") crafted input \citep{Szegedy2013}. Figure \ref{fig:Figure 2} shows an example from image classification.
Adversarial attacks can also interfere with the policies learned by deep reinforcement learning agents \citep{Behzadan2017,Huang2017}.

Adversarial attacks have been shown to be a vulnerability in AI systems, even if attackers do not have access to the underlying model \citep{papernot2017} or if they only have the capacity to modify features in the physical environment \citep{Kurakin2016}. The risks arising from adversarial attacks range from degrading the efficacy of an ML--based decision system (e.g. reducing the reliability of predicted labels on photographs) to undermining security protocols (such as those using facial recognition or biometric images) to interfering with behavior in the real world (such as manipulating the inferences drawn from visual input for an autonomous vehicle). 

Defending against adversarial attacks is a substantial technical challenge. The research literature contains multiple proposed defensive techniques \citep{papernot2017} and demonstrations of how these techniques can fail \citep{Carlini2017,Uesato2018}. But efforts to develop methods for provably robust models appear promising \citep{Katz2017,Singla2019}). The key observation is that the challenge of securing deep neural networks against adversarial attack is one that draws on the same levels of expert analysis that the building of AI systems does. It is therefore a setting in which it is likely that solutions will come from researchers who are engaged in state-of-the-art AI research. 

\subsection{Participants in the market for adversarial robustness regulation} Applying our model to the context of adversarial robustness we first define the relevant actors. 

\subsubsection{Government} We presume a governmental agency that sets the outcome goals for regulation in some domain. The domain might be narrow--the use of specific types of deep learning models in aircraft advisory systems (as in \citet{Katz2017})--or it might be broad--the use of any deep learning models for any purpose in any context. The choice of scope will be in part a function of technical considerations as well as a function of the political environment and the nature of the risks involved. Ensuring against adversarial attack in safety--critical systems that put large numbers of lives at stake may be appropriately allocated to a specialist government agency, such as the FAA in the case of passenger aircraft, with the capacity to regulate along multiple dimensions. In other settings, oversight might be placed in a government agency that specializes in oversight of AI systems across a wide variety of domains. The capacity for oversight of the government agency itself will be a factor: will the electorate, legislative contestants, civil society organizations, and media be better able to evaluate the performance of the agency if it is focused on a familiarly-defined domain, such as airline safety, than if it is focused on a diffuse and novel domain, such as AI safety? And scale will be a consideration: the oversight agency will require sufficient resources to engage in effective oversight but not such immense scale that oversight of the government body itself is diluted. 

We will assume we are looking at deep learning models employed by commercial drones. In this domain, we can imagine a range of solutions in terms of the relevant government oversight bodies. Oversight might be provided by national-level aircraft regulation agencies, like the FAA in the U.S. Or it might be provided by local agencies responsive to the priorities of local communities, much as ride-sharing services are often regulated at the city-level to respond to considerations about transportation, protection of pedestrians and riders, and economic displacement of traditional for-hire car services. 

\begin{figure}[!htb]
    \center{\includegraphics[width=\textwidth]
    {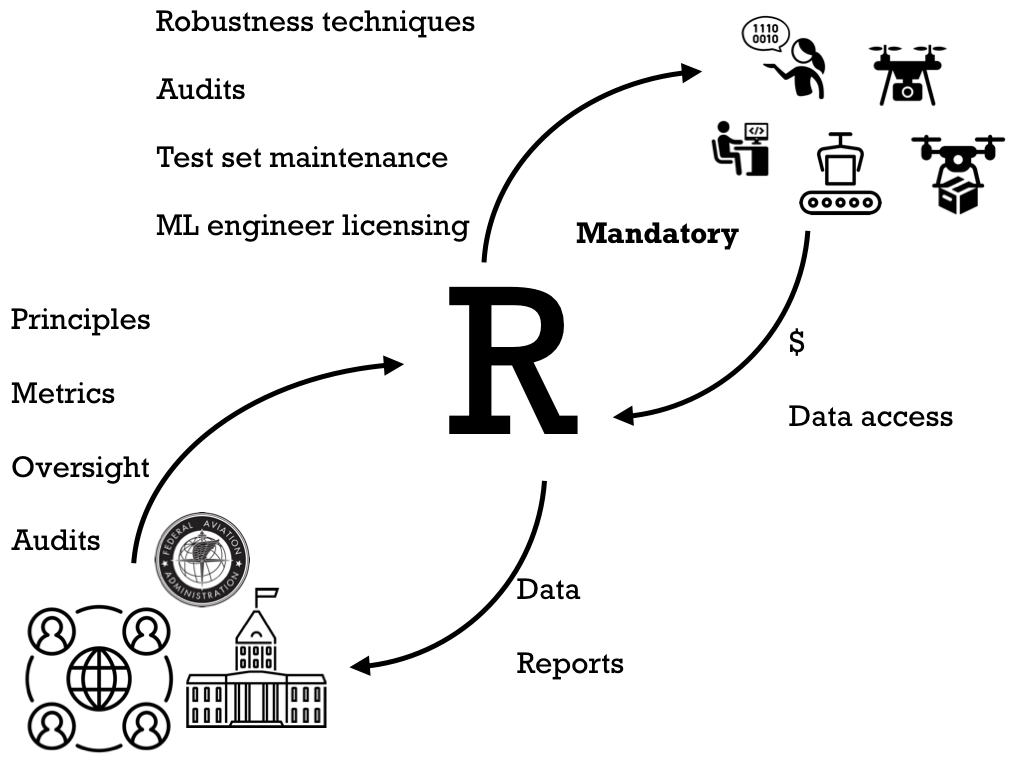}
    \caption{\label{fig:Figure 3} In this market for regulating adversarial robustness in commercial drones, a private regulator (``R'') is licensed to regulate on the basis of outcome metrics and principles supplied by government actors (potentially including existing domestic regulatory agencies and legislatures as well as international bodies.) These government actors would engage in oversight, in part by conducting audits, of the regulator and requiring the regulator to supply data and reports. Drone manufacturers, developers, and companies deploying drones in delivery services would be required by government to purchase the regulatory services of a licensed private regulator, supplying the regulator with data and submitting to robustness regulatory requirements developed and implemented by the regulator, such as auditing, licensing machine-learning engineers and requirements to submit models to accuracy testing against test sets maintained by the regulator prior to deployment.}}
   \end{figure}

\subsubsection{Targets} The targets of regulation are the companies implementing deep learning models in commercial drones and which are required by the government oversight agency to purchase regulatory services from private regulators the agency has approved. Targets might be defined as the companies that sell commercial drones, companies that employ commercial drones (to deliver packages, for example) and/or companies that supply deep learning models (software) to commercial drone manufacturers or users. We can imagine that regulations enacted by the government oversight agency require any entity employing or selling commercial drones to ensure that the deep learning models implemented in the drones have been produced by a manufacturer or developer that is regulated by an approved private regulator. The ultimate target of regulation would need to be capable of producing the data required by and implementing any modifications to the model the regulator. 

\subsubsection{Private regulators} The private regulators here would be private companies (for-profit or non-profit) that invest in the research and development of techniques to evaluate the adversarial robustness of deep learning models in the commercial drone domain and supply regulatory services to targets. Engineers working on technical safety in companies engaged in technical safety research -- for instance, OpenAI, DeepMind, and Google--would hopefully be incentivized to either start or join companies dedicated to providing independent regulatory services. 

\subsection{Licensing} Private regulators would be required to apply for a license to supply adversarial robustness regulatory services in the commercial drone context from the government agency in any jurisdiction in which they wished to operate. If regulation happens at the city level, for example, then a private regulator would apply to individual cities for a license. Cities could conceivably recognize licenses supplied by other cities. For example, if Regulator A is licensed by Los Angeles, San Francisco could accept that license. Similarly, if Regulator B is licensed by San Francisco, Los Angeles might accept that license. Targets wishing to supply or employ commercial drones in either city satisfy their regulatory requirement by purchasing regulatory services from either Regulator A or Regulator B. 

To be licensed to provide regulatory services, a private regulator would demonstrate to the appropriate oversight agency that it satisfies the government’s regulatory criteria. These criteria could include concrete metrics--such as frequency of successful attacks during deployment. But they could also include performance audits, case studies, and reviews. With multiple regulators operating in the market, regulators might be incentivized or required to engage in contests such as those proposed by \citet{Brown2018}. They might also be incentivized to generate metrics that demonstrate safety improvements they have achieved, relative to their competitors. With data from and access to the experiences of multiple regulators, the government oversight agency would be better equipped (than in the case of overseeing a monopoly private regulator) to verify the claims made by regulators about factors such as the efficacy of their systems or the feasibility of improvements. Incentives to advance regulatory standards could be generated either by positive industry reputation spurred by publicity around oversight reports (securing a larger share of the market, particularly where incentives for safety are supported by consumer preferences for safety) or by concrete advantages such as a period of exclusive access to a share of the market.

Government oversight agencies would also be authorized to ensure that the market for private regulatory services in a given domain is competitive. Regulators that are affiliated with targets, for example, will be required to ensure independence and protection of data confidentiality and trade secrets. Total share of the market secured by a given regulator could be capped. Rules could be enacted to ensure that targets can relatively easily switch regulators, such as by requiring regulators to transfer test sets and results for a given target--preserving competitive incentives for regulators to improve the efficiency of their techniques. 

\subsection{Regulatory techniques} The aim of regulatory markets is to create incentives for the private sector to allocate resources--money, talent, and compute--to the challenge of developing more effective methods of defending against adversarial attacks. In addition to investment in techniques that improve robustness--defending a broader class of models against novel adversarial techniques and larger perturbations to input data--we also anticipate that private regulators can supply other services to clients. They could provide expertise on how to do adversarial training and run formal verification on the resulting models during development phases. They could maintain hold-out sets of attack vectors (perturbation types) that they do not show to their clients and against which the client could test their models. 

Regulators would also be competing for clients by trying to improve the efficiency and reduce the cost of robustness. Adversarial training is currently expensive to use. Provided the regulator can continue to demonstrate that its techniques meet oversight requirements, it would face an incentive to develop lower-cost training algorithms. Adversarial training can also come with an efficiency loss--making commercial drones less effective at performing the tasks sought by users. Regulators would also be competing to develop robustness techniques with lower efficiency costs. 

We envision that the license awarded for a private regulator might initially be narrow in scope, permitting, for example, only regulatory oversight of the adversarial vulnerability of deep learning models of a particular class in commercial drones. But a virtue of creating a market for regulation is that private regulators will have an incentive to expand their market access. This would play out through proposals made by regulators to government oversight agencies for an expanded license. Regulators might seek to develop techniques to verify robustness in a broader set of deep learning models. They might seek to expand their remit beyond implementations in commercial drones. Or they might seek to expand their regulatory capacity to other dimensions of the commercial drone business: developing standards for safe exploration by drones, for example.

\subsection{Transitioning to regulatory markets} We have sketched this prototype under the assumption that government oversight agencies might create regulatory markets from the top down. This may not be too much to expect in the domain we are considering--commercial drones--because this is a novel domain and there are few pre-existing regulatory regimes. Experimenting with a new approach seems possible here. Teams in industry and academia are already focused on this area of research and in this limited domain sufficient safeguards for data and independence could be put in place to enable existing safety teams within established organizations to experiment with providing regulatory services to users outside of their organization. 

A government that was ready to prototype regulatory markets for AI regulation in this highly specified domain would need to undertake the following steps. First, a government oversight body would need to be established. This could be established within an existing regulatory agency, such as the FAA in the U.S. But we suspect, particularly in light of what has been learned about oversight efforts at the FAA recounted in our discussion of the Boeing 737 MAX crashes, the effort of regulatory innovation may be better served by establishing a novel, targeted, overseer, specifically tasked with implementing outcomes-based and risk-based regulatory methods. Second, enabling legislation would need to be enacted that creates the private regulatory regime: establishing outcome metrics or principles or, more likely, authorizing the overseer to set metrics or outcomes based on legislated principles to protect against adversarial attacks in commercial drone settings; authorizing the overseer to license and supervise private regulators based on their ability to meet and maintain these outcomes; and requiring commercial drone manufacturers, developers, and users to purchase the regulatory services of a licensed regulator.  This legislation likely would also have to preempt other regulatory requirements imposed on regulatory targets in this domain.  It would be important for the oversight body to be tasked, especially initially, not only with implementing the private regulatory regime but also with collecting and monitoring data on the performance of the regulators and targets to ensure that regulation is effective and the market is adequately competitive. The oversight body would also require either the authority to directly supplement the enabling legislation to impose any requirements needed to promote efficacy and competition or the capacity to propose legislative changes to the appropriate legislative body. We can imagine that a government interested in prototyping this approach could engage in a carefully-reviewed pilot to develop final versions of enabling legislation and regulatory/market design.

We can also imagine, however, that in the absence of legislative initiative, this novel regulatory approach might grow, bottom up, out of industry self-regulatory efforts. Much regulation in technically complex or novel areas originates in such efforts. Here, companies involved in the development of commercial drones--developers of deep learning models to be deployed in drones, drone manufacturers, companies anticipating large-scale deployment of drones--might, in recognition of the risks of adversarial attack, organize as a consortium to function, initially, in the place of a government oversight agency. The consortium would establish oversight criteria, metrics and procedures. It would then invite entities--including internal units of consortium members, operating under rules for independence and confidentiality--to apply for authorization to regulate. These regulators would develop robustness techniques and procedures for certifying the robustness of deep learning models and implementations. Public certifications could create an incentive for drone developers and suppliers to obtain certification. The consortium could commit to developing criteria for oversight that anticipate the kinds of oversight that a legitimate public agency could require and developing a transition plan for oversight to be eventually handed to a public agency once proof-of-concept has been completed. 

\section{Conclusion}
Building safe machine learning systems requires not only technical innovations, it also requires regulatory innovations.  Research is advancing on techniques for certifying adversarial robustness, for example \citep{Katz2017,Singla2019}. The question remains, however, how actors will be required or incentivized to implement safety techniques or submit to certification and how certification will be conducted. Conventional approaches to regulation through government agencies may work, but there are many doubts about the capacity of governments to regulate on the scale and in the time-frame of rapid and complex AI innovation. Self-regulation may also succeed, but there are obstacles there too--both in terms of reliability and in terms of legitimacy. Regulatory markets offer the potential to harness the speed and complexity benefits of private markets, while ensuring the legitimacy of regulation, all in the service of safer AI systems. We are not naive about the challenges to be overcome; regulatory markets will require careful design and robust government oversight. They will not be appropriate in all contexts. But the challenges here seem surmountable, in ways that the responding to the failures of traditional regulation in complex technology markets do not. At a minimum we urge governments and industry to begin to explore this new model for regulation as a possible response to the mounting urgency of reigning in the risks of powerful AI.



\bibliography{iclr2019_conference}
\bibliographystyle{iclr2019_conference}

\end{document}